\documentclass{article}
\usepackage{INTERSPEECH2022}
\usepackage{amsmath,graphicx}
\usepackage{xcolor}
\usepackage{url}
\usepackage{hyperref}
\usepackage{balance} 

\def\L{{\cal L}}
\def\R{{L_{\mathrm{rec}}}}

\def\Rev{{\cal R}}
\def\Dry{{\cal D}}

\title{CycleGAN-Based Unpaired Speech Dereverberation}
\name{Hannah Muckenhirn$^1$, Aleksandr Safin$^2$, Hakan Erdogan$^1$, F\'elix de~Chaumont~Quitry$^1$, Marco Tagliasacchi$^1$, Scott Wisdom$^1$,  John R. Hershey$^1$}
\address{$^1$Google Research\\
  $^2$Skolkovo Institute of Science and Technology}
\email{muckenhirn@google.com, aleksandr.safin@skoltech.ru, \{hakanerdogan, fcq, mtagliasacchi, scottwisdom, johnhershey\}@google.com}
\begin{document}
\maketitle
\begin{abstract}
Typically, neural network-based speech dereverberation models are trained on paired data, composed of a dry utterance and its corresponding reverberant utterance. The main limitation of this approach is that such models can only be trained on large amounts of data and a variety of room impulse responses when the data is synthetically reverberated, since acquiring real paired data is costly.
In this paper we propose a CycleGAN-based approach that enables dereverberation models to be trained on unpaired data. We quantify the impact of using unpaired data by comparing the proposed unpaired model to a paired model with the same architecture and trained on the paired version of the same dataset. We show that the performance of the unpaired model is comparable to the performance of the paired model on two different datasets, according to objective evaluation metrics. Furthermore, we run two subjective evaluations and show that both models achieve comparable subjective quality on the AMI dataset, which was not seen during training.
\end{abstract}
\noindent\textbf{Index Terms}: Dereverberation, unpaired training, CycleGAN
\section{Introduction}
\label{sec:intro}
In speech processing and communication applications,  reverberation can reduce intelligibility and signal quality~\cite{naylor2010speech}.
Reverberation occurs when sound is reflected off surfaces, such as walls, ceilings, and furniture, adding delayed versions of the sound that, in effect, blur the signal across time.  This effect is more pronounced in large rooms with reflective surfaces when the microphone is far away from the speech source.  

The goal of \emph{dereverberation} is to remove reverberation while preserving the non-reverberated (\emph{dry}) signal. In this paper, we focus on single channel speech dereverberation. In recent years, there has been a shift from signal processing-based methods, such as \textit{weighted prediction error} (WPE)~\cite{WPE:nakatani_2010}, to neural network (NN) based methods for dereverberation. As such, DNN-WPE~\cite{DNN-WPE:kinoshita_2017} uses a deep neural network to estimate the power spectrum of the target signal for WPE. Wang et al.~\cite{TC:wang_2020} use a convolutional neural network (CNN) to estimate real and imaginary parts of dry speech directly from the reverberant one. 

Such NN-based models have been trained using pairs of aligned audio clips of dry and reverberated data, either with supervised training \cite{xiao2016speech, luo2018real} or generative adversarial networks (GANs) \cite{su2020hifi, su2021hifi}. The model takes as input a reverberated audio clip and is trained to produce the corresponding dry audio clip. 

Paired data can be obtained in two ways. The first is to record the same speech with a pair of microphones: one  close to the speaker's mouth, which records less reverberant speech, and the other farther away from the speaker, which records more reverberant speech. The AMI dataset~\cite{AMI:mccowan_2005}, for example, was recorded this way. Due to the special recording setup, the amount and variety of paired data of this type is limited.
It is also difficult to achieve a strong contrast in reverberation without introducing such noises as breaths and lip smacks in the near microphone or sensor and environment noise in the far microphone.   
Other problems include time misalignment and differences in frequency response between the two microphones. These spurious inter-microphone differences can interfere with effective learning of dereverberation.  
    
The second way to obtain paired data is to generate synthetic reverberated signals either through simulated or measured room impulse responses~\cite{DNN-WPE:kinoshita_2017, TC:wang_2020, REVERB:kinoshita_2016}. This has the advantage that one can use large amounts of dry speech as a source and synthesize the corresponding reverberated audio clips.
However, the availability of measured room impulse responses is limited and may not cover all scenarios in the target domain. 
Measured room impulse responses are also linear and time-invariant, whereas real scenarios can have nonlinear and time-varying acoustic phenomena, such as motion effects.   Using simulated room impulse responses addresses these problems by covering a larger space of possible room characteristics, and potentially allowing simulation of motion effects and other phenomena.  However, there are many ways in which synthetic reverberation may fail to match real reverberation, and it is challenging to close this gap.  All of these factors may impair generalization from synthetic reverberation to real-world reverberation. 

This paper addresses these problems by training dereverberation models  on unpaired data, using large independent collections of dry speech and real reverberant speech data.
To train such models we use a cycle-consistency loss for unpaired training inspired by CycleGAN~\cite{CycleGAN:zhu_2017}, which was developed for unpaired image-to-image translation.
CycleGANs have previously been used in unpaired speech denoising~\cite{CSE:meng_2018, AIA-CycleGAN:yu_2021, MO-CycleGAN:xiang_2021}.   CycleGANs have been used on log Mel filterbank features to improve downstream tasks such as automatic speech recognition \cite{mimura2017cross} and speaker verification \cite{nidadavolu2020single} for noisy and reverberant speech.  Voice conversion has also been performed using CycleGANs~\cite{CycleGAN-VC:kaneko_2018}.  However this is, to the best of our knowledge, the first time that such an approach is proposed for dereverberation. 

In this paper, we show that unpaired training is effective for the task of dereverberation. To do so, we compare the performance of our proposed unpaired model with a paired model that has the same architecture and was trained on a paired version of the same dataset. This enables us to quantify the impact of using unpaired data only, without having other parameters to take into account, such as different architecture or training data. We show via objective and subjective evaluation that our unpaired model performs comparably to the paired model, especially on out-of-domain data.   Although we focus on synthetic data in this paper, validation of the unpaired approach is a key step in enabling the use of large amounts of real training data.

\section{Proposed approach}
\label{sec:approach}
In this section, we describe our proposed approach, which is based on CycleGAN~\cite{CycleGAN:zhu_2017}. 
We use the following notations. Let $\Rev$ denote a set containing reverberant speech and $\Dry$ denote a set containing dry speech, representing the source and target distributions for dereverberation.
In terms of content, speaker identities, emotional content, gender, accent, we assume the sets are balanced and mostly equivalent.

These sets can be prepared in various ways. One option is to select real recordings based on their reverberation properties, such as T60 and direct-to-reverberation ratio.  The reverberation level may be known or may be estimated using a model (e.g., \cite{lim2015acoustic, gamper2018blind, bryan2020impulse,slaney2022}).
In this paper, we take the approach of generating the reverberated dataset from a relatively dry speech set.
This produces data that can be treated as either paired or unpaired, allowing us to evaluate and directly compare these two approaches.

During training, the model is presented with ``unpaired'' tuples $(x_R,y_D)$, where $x_R$ is reverberated and $y_D$ is dry. As stated before, $x_R \in \Rev$ and $y_D \in \Dry$ are not paired, e.g., they have different speech content and come from different speakers. We do not have access to $x_D$, i.e., the dry version of $x_R$, nor to $y_R$, i.e., the reverberated version of $y_D$.

Our model contains the following components:

\noindent \emph{Generator $G_{R\rightarrow D}$}: removes reverberation and produces dry speech.

\noindent \emph{Generator $G_{D\rightarrow R}$}: adds reverberation and produces reverberant speech.

\noindent \emph{Discriminator $D_R$}: distinguishes real reverberant speech from the outputs of generator $G_{D\rightarrow R}$.

\noindent \emph{Discriminator $D_D$}: distinguishes real dry speech from the outputs of generator $G_{R\rightarrow D}$.

The generators $G_{R\rightarrow D}$ and  $G_{D\rightarrow R}$ are trained jointly using a mix of seven losses: $\L_{G_{R\rightarrow D}}$, $\L_{G_{D\rightarrow R}}$, $\L_{\mathrm{cycle}_R}$, $\L_{\mathrm{cycle}_D}$, $\L_{\mathrm{feat\_cycle}_R}$, $\L_{\mathrm{feat\_cycle}_D}$, and $\L_{id_D}$, further detailed below.

$\L_{G_{R\rightarrow D}}$ and $\L_{G_{D\rightarrow R}}$ correspond to the adversarial losses, adopting the following hinge-loss variant:
\begin{equation} \label{eq:G_R2D_loss}
    \L_{G_{R\rightarrow D}} = E_{x_R}[\max(1-D_D(G_{R\rightarrow D}(x_R)],
\end{equation}
\begin{equation}
    \L_{G_{D\rightarrow R}} = E_{y_D}[\max(1-D_R(G_{D\rightarrow R}(y_D)],
\end{equation}
where $E_{x_R}$ and $E_{y_D}$ indicate expectations.

We define cycle outputs $\tilde x_R = G_{D\rightarrow R}\left(G_{R\rightarrow D}(x_R)\right)$ and $\tilde y_D = G_{R\rightarrow D}\left(G_{D\rightarrow R}(y_D)\right)$, used to compute the cycle losses:
\begin{equation}
    \L_{\mathrm{cycle}_R} = E_{x_R}[\R\left(\tilde x_R, x_R\right)],
\end{equation}
\begin{equation}
    \L_{\mathrm{cycle}_D} = E_{y_D}[\R\left(\tilde y_D, y_D\right)],
\end{equation}
with $\R$ a multi-scale spectrogram reconstruction loss \cite{su2021hifi, engel2020ddsp}.

In addition, we have a feature loss on each discriminator's intermediate layer activations as follows:
\begin{equation}
\L_{\mathrm{feat\_cycle}_R} = E_{x_R}\left[\sum_l\left| D_R^{(l)}(\tilde x_R) - D_R^{(l)}(x_R)\right|\right],
\end{equation}
\begin{equation}
\L_{\mathrm{feat\_cycle}_D} = E_{y_D}\left[\sum_l\left| D_D^{(l)}(\tilde y_D) - D_D^{(l)}(y_D)\right|\right].
\end{equation}
Here $D_R^{(l)}(x)$ and $D_R^{(l)}(x)$ indicate the $l$th layer activation of the reverberant and dry discriminators respectively.

Finally, we provide an identity loss which enforces the equality of output to the input when using the dereverberation model on an already dry signal:
\begin{equation}
    \L_{\mathrm{id}_D} =  E_{y_D}[\R\left(G_{R\rightarrow D}(y_D), y_D\right)].
\end{equation}
Note that we do not use an identity loss for the generator $G_{D\rightarrow R}$, since this generator always adds reverberation, even when the input is already reverberated.

Overall, the loss of the generators can be written as:
\begin{equation} \label{eq:G_total_loss}
\begin{aligned}
    \L_{G} = & \lambda_{\mathrm{gan}} \left(\L_{G_{R\rightarrow D}} + \L_{G_{D\rightarrow R}}\right) + \lambda_{\mathrm{cycle}}(\L_{\mathrm{cycle}_R} + \L_{\mathrm{cycle}_D})\\
   & + \lambda_{\mathrm{feat}} (\L_{\mathrm{feat\_cycle}_R} + \L_{\mathrm{feat\_cycle}_D})
    +  \lambda_{\mathrm{id}} \L_{\mathrm{id}_D} 
\end{aligned}
\end{equation}
The losses used to train the generator serve different purposes and finding the right balance among them can be challenging. The goal of $\L_{G_{R\rightarrow D}}$ is to ensure that the outputs of $G_{R\rightarrow D}$ are dry. The goal of $\L_{G_{D\rightarrow R}}$ is to ensure that the outputs of $G_{D\rightarrow R}$ are reverberated. The purposes of $\L_{\mathrm{cycle}_R}$, $\L_{\mathrm{cycle}_D}$, $\L_{\mathrm{feat\_cycle}_R}$, $\L_{\mathrm{feat\_cycle}_D}$ and  $\L_{id_D}$ is to ensure that the speech characteristics, such as phonetic content and speaker identity, are not modified by the generator. We have seen in practice that when $\lambda_{\mathrm{id}}$, $\lambda_{\mathrm{cycle}}$ or $ \lambda_{\mathrm{feat}}$ are set to 0, the model either diverges or its performance is significantly impacted. On the other hand, if $\lambda_{\mathrm{id}}$, $\lambda_{\mathrm{cycle}}$ or $ \lambda_{\mathrm{feat}}$ are too large, the generators learn an identity mapping.

The dry discriminator $D_D$ is trained with the following GAN discriminator loss:
\begin{eqnarray} \label{eq:D_D_loss}
    \L_{D_D} & = & E_{y_D}[\max(1 - D_D(y_D)] + \nonumber \\
               & & E_{x_R}[\max(1 + D_D\left(G_{R\rightarrow D}\left(x_R\right)\right)].
\end{eqnarray}
The reverberant signal discriminator's loss is defined similarly. The total discriminator loss is simply
$\L_{D} = \L_{D_D} + \L_{D_R}.$

\section{Architecture}
\label{sec:architecture}
\begin{figure}[t]
  \centering
  \includegraphics[width=0.73\linewidth]{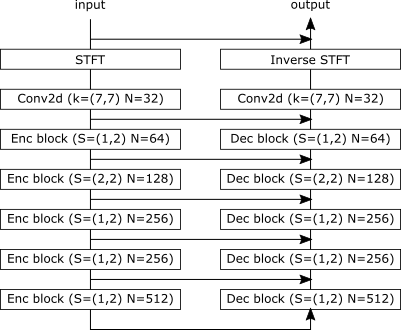}
  \vspace{-5pt}
  \caption{Architecture of the generators.}
  \label{fig:archi_generators}
  \vspace{-5pt}
\end{figure}
In this section, we describe the architecture of the generators $G_{R\rightarrow D}$ and $G_{D\rightarrow R}$ and the discriminators $D_R$ and $D_D$.

The generators $G_{R\rightarrow D}$ and $G_{D\rightarrow R}$ have exactly the same architecture. They take as input a complex short-time Fourier transform~(STFT), computed on the input waveforms with a window length of 20 ms (320 samples) and an overlap of 10 ms. The complex STFT is fed to the generators as a two-channel image, where the two channels are the real and imaginary parts of the STFT. The generators are 2D-convolutional UNets, which follows the 
architecture described in~\cite{STFT-SEANet:li_2020}. The UNet is composed of an encoder and a decoder, which is a mirrored version of the encoder. There are skip connections between each encoder block and the corresponding decoder block.

Each encoder block is composed of two convolutional layers. The first one has a kernel size of 3x3 and a stride of 1x1. The second one has either a kernel size of 3x4 and a stride of 1x2 or a kernel size of 4x4 and stride of 2x2. The overall architecture is illustrated in Figure~\ref{fig:archi_generators}.

The discriminators $D_R$ and $D_D$ have the same architecture as MelGAN~\cite{kumar2019melgan}. 
Each discriminator is multi-scale: it is composed of three discriminators that have the same architecture and take as input a raw waveform that is down-sampled respectively by x1, x2 and x4. Each discriminator is fully convolutional and is composed of the following layers: 1 convolution, 4 grouped convolutions and 2 convolutions.

\begin{table*}
    \caption{
        Evaluation of paired and unpaired models on the evaluation subset of the Librivox dataset and the AMI dataset. For FWSegSNR and SDR, higher values mean better, for estimated T60, lower is better.
    }
    \label{tab:eval_on_filtered_librivox_ami}
    \vspace{-5pt}
    \centering
    \begin{tabular}{|l|c|c|c||c|c|c|}\hline
    & \multicolumn{3}{c||}{Librivox} &  \multicolumn{3}{c|}{AMI} \\ \hline
    Models & FWSegSNR & SDR & Estimated T60 & FWSegSNR & SDR & Estimated T60 \\
    \hline
    Paired model & $15.5 \pm 0.04$ & $14.2 \pm 0.06$  & $0.10 \pm 2\times 10^{-4}$& $3.3\pm 0.07$ & $2.5\pm 0.15$ & $0.12\pm 6\times 10^{-4}$\\
    Unpaired model & $15.5 \pm 0.06$ & $12.8 \pm 0.06$  & $0.11 \pm 4\times 10^{-4}$ & $3.2\pm 0.07$ & $1.6\pm 0.14$ & $0.12\pm 9\times 10^{-4}$\\
    No model & $13.1 \pm 0.10$ & $11.4 \pm 0.08$ & $0.38\pm 3\times 10^{-3}$ & $2.6\pm 0.07$ & $1.6\pm 0.14$ & $0.25\pm 3\times 10^{-3}$  \\
        \hline
        \end{tabular}
\end{table*}

\begin{table}
    \caption{
        Results of subjective evaluation. Percentage of votes for each model on two criteria: which model produces less reverberation and which model produces better quality audio. The p-values are $<10^{-3}$ on Librivox, and 0.45 and 0.15 for ``less reverb'' and ``better quality'', respectively, on AMI.
    }
    \label{tab:subjective_eval}
    \vspace{-5pt}
    \centering
    \begin{tabular}{|l|c|c||c|c|}\hline
    & \multicolumn{2}{c||}{Librivox} &  \multicolumn{2}{c|}{AMI} \\ \hline
    & Paired & Unpaired & Paired & Unpaired \\
    \hline
    Less reverb & 61.5\% & 38.5\% &  51.2\% & 48.8\% \\
    Better quality & 59.5\% & 40.5\% & 52.3\% & 47.7\% \\\hline
    \end{tabular}
\end{table}

\section{Experimental setup}
\label{sec:experiments}

\subsection{Datasets}
\label{ssec:datasets}
\vspace{-1mm}
The dereverberation models are trained on synthetically reverberated data. The dry audio samples were taken from the Librivox dataset~\cite{librivox}, which contains audio books. Utterances that were  shorter than 1 second were removed and a speech activity detector was used to extract speech segments. The data and its processing is described in~\cite{hoover2017putting}. For our use case, we additionally filter the data to remove reverberant and/or noisy utterances. To do so, we use a criterion based on a speech enhancement model with the same architecture as the generator, trained to perform dereverberation and denoising. If the output of this model is significantly different from the input, then we discard the audio clips as it means that it contains noise or/and reverberation. This is done by computing the ratio of the standard deviation of the output and the standard deviation of the residual, computed over rolling windows of 3 seconds. Two criteria are used: 1) if any window has a ratio smaller than a chosen threshold or 2) if the logarithm mean is smaller than a chosen threshold, then the entire audio clip is removed. This process removes around 40\% of the data. We then split the audio clips into 3 second-long clips. We finally split the remaining data into two sets: the training set, which contains 18 million utterances and the evaluation set, which contains 10000 utterances.

From these dry utterances, we synthetically generate paired reverberated utterances on-the-fly by convolving them with room impulse responses (RIRs) randomly chosen from a large set of simulated RIRs. We simulated 650,000 and 30,000 room impulse responses for train and test sets using the image method with frequency-dependent wall reflections. The room dimensions are uniformly randomly chosen between 3-7 meters in width, 4-8 meters in length and 2.13-3.05 meters in height. We randomly choose among 5 wall materials, 4 floor materials and 4 ceiling materials with varying frequency dependent acoustic reflection characteristics. When applying the image method, we artificially jitter the image locations randomly inside a cube with a side length of 16 cm to avoid the sweeping echo phenomenon \cite{Sena2015OnTM}. Our ``dry'' targets actually include the early reverberation part of an RIR, as we have seen in practice that it improves the performance of the dereverberation models. To do so, we apply a rectangular window of 20ms width to each RIR, i.e. the values of the RIR 20 ms after its peak value are set to 0, and we convolve it with the dry source data.

The unpaired model sees two different sets of data during its training and learns to transform the data from one set to another: a set composed of dry utterances and a set of reverberant utterances. Thus, we need to ensure that all the data in the first set is dry and all the data in the second set is reverberant. The first case is ensured by our filtering process of the Librivox dataset. We ensure the second case by only using RIRs that produces reverberant ``enough'' signal, by filtering out RIRs with a reverberation time T60 lower than 400ms. The remaining 127,000 RIRs have T60 values $\in [0.4, 1.2]$ seconds and DRR values $\in [-19.0, 18.3]$ dB. Note that these constraints do not apply to the paired models, as the paired data is always ``ordered'', i.e. the reverberant utterance is always equally or more reverberated than its corresponding dry utterance. However, we use the same data for the paired model to have a fair comparison.

Finally, to simulate unpaired data, we simply split the training data into two halves and used the dry utterances from the first half and the reverberated utterances from the second half. This ensures that the model never sees the same utterance in its two different versions: dry and reverberated. 

We evaluate the dereverberation models on two paired datasets. The first one is the evaluation set that corresponds to the data used for training: the dry data was taken from the evaluation set of the filtered Librivox dataset and the reverberant paired data is synthetically generated from it. It is worth noting that a different set of room impulse responses were generated so that the rooms were not already seen during the training.
We also evaluate the models on real reverberant data. We use recordings from the AMI dataset~\cite{AMI:mccowan_2005}, which consists of meetings recorded in 3 different rooms. Each participant was equipped with a headset microphone, in addition to the room far field microphones. To compute the evaluation metrics, we use the headset recordings as targets and the far-field microphone recordings as reverberant inputs. It is worth noting that the far field data is not only reverberant, it is also noisier than the target and misaligned. The headset recordings also have breathing noise that the far field microphones do not capture. The estimated T60 values of Librivox and of AMI are respectively comprised in the range $[0.17, 0.98]$ and $[0.12, 0.64]$. 
\vspace{-1mm}
\subsection{Metrics}
\vspace{-1mm}
To evaluate our models, we use 3 metrics: signal-to-distortion ratio (SDR)~\cite{fevotte2005bss_eval}, frequency-weighted segmental SNR (FWSegSNR)~\cite{tribolet1978study, hu2007evaluation} and an estimation of the reverberation time (T60). We report SDR instead of the more commonly used SI-SDR metric, because the AMI dataset is misaligned. SDR allows a misalignment of up to 512 samples. FWSegSNR is commonly used for evaluating dereverberation models, it was for example used in the REVERB challenge~\cite{REVERB:kinoshita_2016}. For the third metric, we use a neural network-based model that was trained on synthetic reverberant and noisy data to predict T60 values, described in ~\cite{slaney2022}. T60 is the time it takes for a reverberation to decay by 60 dB.

\vspace{-1mm}
\subsection{Paired model}
\vspace{-1mm}
In Section~\ref{sec:results}, we will compare our proposed unpaired model to the equivalent paired model. The paired model is composed of one generator and one discriminator, which correspond to $G_{R\rightarrow D}$ and $D_D$, with the same architecture described in Section~\ref{sec:architecture}. The generator is trained with the loss $\L_{G_{R\rightarrow D}}$ described in (\ref{eq:G_R2D_loss}) and a paired feature loss~\cite{kumar2019melgan,zeghidour2021}, computed between the output and paired target:
$\L_{G} = \L_{G_{R\rightarrow D}} + 100 \times \L_{\mathrm{feat}}.$
The  discriminator is trained with the loss $\L_{D_D}$ described in (\ref{eq:D_D_loss}).

The paired model is trained on the same dataset as the unpaired model, described in section \ref{ssec:datasets}. However, it sees paired pairs of utterances $(x_D, x_R)$, where $x_R$ is the synthetically reverberated version of $x_D$.

\vspace{-1mm}
\subsection{Training}
\vspace{-1mm}
Both models were trained with a batch size of 32 and a learning rate of 0.0001 for the generators and 0.001 for the discriminators. Both models are trained on input of 512 ms with a sampling rate of 16 kHz.
Each input is peak normalized and a uniform random gain between 0.3 and 1.0 is applied.
For the unpaired model, we used the following loss weights: $\lambda_{\mathrm{gan}} = 1.0, \lambda_{\mathrm{cycle}} = 0.1, \lambda_{\mathrm{feat}} = 1.0,
\lambda_{\mathrm{id}} = 0.5$.

\section{Results}
\label{sec:results}
In this section we present the results obtained with the paired model and the proposed unpaired model. Paired model performance serves as an upper bound on the unpaired model's performance, since the paired model has access to more information.

In Table~\ref{tab:eval_on_filtered_librivox_ami}, we present the performance of the two models, evaluated on the test set of Librivox as well as on the test set of the AMI dataset, which was not seen during training and is composed of real reverberant audio clips. In the row ``No model'', we present the evaluation metrics computed on the reverberant input data, not processed by any model. For each metric, we also provide 95\% confidence intervals, obtained with non-parametric bootstrapping. We observe that the FWSegSNR and estimated T60 metrics of the two models are on par on the two datasets, indicating that the unpaired model reaches a similar level of performance as the paired model. On the other hand, SDR is lower for the unpaired model, by 10\% on Librivox and by 36\% on AMI dataset. SDR, while allowing for a small misalignment, is a sample-by-sample metric. On the other hand FWSegSNR is computed on a frame-level and the estimated T60 is computed on an utterance-level. One possible explanation for this discrepancy between SDR and the two other metrics is that the unpaired model does not have any incentive to produce perfectly aligned outputs on a sample level, since it never sees aligned or paired data.

To confirm the results obtained with the objective evaluation metrics, we also present the results of two subjective evaluations. The goal of the first evaluation was to compare the amount of reverberation in the outputs of the paired and unpaired models, while the goal of the second was to compare the overall audio quality. Both evaluations were run in the form of A/B tests. In each question, the raters were presented with two audio clips, corresponding respectively to the outputs of the paired and unpaired models, run on the same reverberant audio clip. They were asked to choose which audio clip had the smallest amount of reverberation (for the first task) and which audio clip had the best overall audio quality (for the second task). The order in which the audio clips were presented was shuffled for each question as to not advantage one model over the other.
The reverberant audio clips were taken from the same datasets used to compute the evaluation metrics: 100 utterances from the filtered librivox dataset (synthetic reverberation), and 100 utterances from the AMI dataset (real reverberation). All audio clips have a duration of  3 to 5 seconds. Each question was answers by ten raters.

The results are presented in Table~\ref{tab:subjective_eval}. We observe that on Librivox, the paired model gets $\approx$60\% of the votes for both tasks. On the other hand, both paired and unpaired models get approximately the same number of votes on the AMI dataset. To confirm these observations, we performed a Wilcoxon signed-rank test and present the corresponding p-values. On Librivox, the p-values are lower than $10^{-3}$ for both rating tasks, which means that we can reject the null hypothesis that both models have the same distribution of votes. On the other hand, on AMI, the p-values are respectively equal to 0.45 and 0.15 for both tasks, which means that we cannot reject the null hypothesis. The fact that the paired model is somewhat preferred over the unpaired model on Librivox is expected as it was trained with more information about this specific dataset and specific type of reverberation. On the other hand, both models generalize equally well to unseen data and an unseen type of reverberation. This shows that the unpaired models is capable of performing as well as the same model trained in a paired manner \footnote{Audio examples available at \url{https://google-research.github.io/seanet/unpaired-dereverb/examples/}}.

\section{Conclusion}
\label{sec:conclusion}
We proposed an approach that enables speech dereverberation models to be trained with unpaired data. We showed that the proposed unpaired model performs comparably to an equivalent paired model, both on in-domain and out-of-domain data, on two evaluation metrics: FWSegSNR and estimated T60 values, which are computed on a frame and utterance level. On the other hand, the sample-level SDR metric favors the paired model, which is likely due to the fact that the unpaired model has no incentive to match perfectly the target signal on a sample level. We also conducted two subjective evaluations, which revealed that the paired model overfits better to in-domain data but both models perform comparably on out-of-domain data. Given that unpaired models compare favorably to paired models,

One thing to note in the presented work is that we treated the problem of unpaired dereverberation as a  one-to-one mapping problem. However, in the case of reverberation, the generator that adds reverberation would correspond to a one-to-many mapping, because the amount of reverberation can vary. In future work, we plan to exploit this and add a latent encoder, which will encode reverberation information and will serve as conditioning to transform the one-to-many mapping into a one-to-one mapping problem. This latent encoder could either be trained in an unsupervised manner, e.g. with a contrastive loss, or trained in a supervised manner, e.g. to predict T60 values.

\vfill\pagebreak

\bibliographystyle{IEEEtran}
\balance
\bibliography{refs}

\end{document}